# Origin of hysteretic field splitting of the Zero Bias Conductance Peak in (100) and (110) oriented $Y_1Ba_2Cu_3O_{7-x}$ films

R. Beck, A. Kohen, G. Leibovitch, H. Castro and G. Deutscher

School of Physics and Astronomy, Raymond and Beverly Sackler Faculty of Exact Science, Tel Aviv University Ramat Aviv 69978, Israel

*We have studied the evolution of the Zero-Bias Conductance Peak (ZBCP) splitting under applied magnetic fields in tunneling experiments on $Y_1Ba_2Cu_3O_{7-x}$(YBCO), and particular its hysteresis. We have been able to distinguish between two possible contributions to the splitting. One of them is connected to Meissner screening currents [1] whose variation in increasing fields is governed by the Bean-Livingston barrier [2] that delays flux entry well above the lower thermodynamical critical field $Hc_1$, up to the fields of the order of the thermodynamical critical field Hc. The other contribution, dominant in (110) oriented films, is seen in decreasing fields where there are no Meissner screening currents, since there is no barrier to flux exit and it may be connected to the magnetic induction in the sample as proposed by Laughlin [3].*

PACS numbers: 74.50.+r , 74.72.Bk

## 1. INTRODUCTION

The quest for the symmetry of the order parameter (OP) in High-Tc Superconductors has been pursued intensively, due to the potential to reveal the mechanism responsible for superconductivity. It has been already well established that the dominant part of the OP is a d-wave symmetry, although the existence of a second sub dominant OP is still under debate. One main fingerprint of the d-wave symmetry is the ZBCP, regularly seen in in-plane tunneling experiments, originated by the Andreev Surface Bound States (ABS). Those states are created as a result of a phase change by $\pi$ in the OP after reflection from a (110) surface in a d-wave superconductors [4].The existence of a ZBCP in other than (110) oriented films is attributed to surface roughness [1]. This ZBCP is split in the present of a magnetic field applied parallel to surface of the sample, as was first noticed by Lesueur et al. [5]. A possible explanation of the splitting due to magnetic impurities is ruled out since the splitting is highly anisotropic in a way that it is maximum when applying the magnetic field parallel to the CuO plane [6,7].

Another interpretation of the ZBCP splitting was then proposed by Fogelstrom et al. (FSR), in which the splitting is caused by a Doppler shift of the condensate momentum due to Meissner screening currents. Also, as previously shown [6,7,8], the splitting is highly hysteretic, i.e. for a given field the splitting in increasing fields in much larger than in decreasing ones. This behavior cannot be explained by the FSR model alone. Our proposal is to take into account the existence of a Bean-Livingston barrier together with the FSR model in order to explain that hysteretic behavior. This Bean Livingston barrier is a direct outcome of the attraction between first vortex entry and an image antivortex near to the superconducting surface, and an attraction of the vortex to the interior of the sample due to Lorenz force with the screening currents. The effect of this barrier is to delay flux entry (e.g. in increasing fields) well above the lower thermodynamical critical field $Hc_1$, up to the fields of the order of the thermodynamical critical field Hc. In contrast the barrier effectively do not exists for flux exit (e.g. in decreasing fields), as seen in other tunneling experiments [9].

While explaining the existence of the hysteresis, this cannot explain all experimental results for (110) oriented films, which need to be explained by another origin of the ZBCP splitting in that orientation. That can be the appearance of an additional sub dominant OP as proposed by Laughlin [3].

## 2. EXPERIMENTAL

We have repeated the tunneling experiments for (110) and (100) oriented YBCO films. We have grown the two in-plane orientation in same time, this way the doping and other external conditions on both samples should be identical. The 3200A thick YBCO samples having a down set critical temperature of 89K, were grown on (110) SrTiO3 and (100) LaSrGaO$_4$ substrates respectively, using rf and dc, off-axis sputtering,

## Origin of hysteretic field splitting of the ZBCP in (100) and (110) oriented $Y_1Ba_2Cu_3O_{7-x}$ films

based on the method described in reference 10. SEM, AFM and X-ray diffraction tests were used to confirm the desired orientation of the samples with a good oriented morphology. In-plane orientation was also confirmed by the strong in plane anisotropy of the in-plane normal state conduction. Surface roughness was of the order of 100Å. Tunneling junctions were formed by pressing Indium contacts onto the surface of freshly prepared samples [6,11]. Magnetic field was aligned parallel to the film and to the CuO planes during the experiment.

### 3. RESULTS

While the zero field dI/dV characteristics of the (100) and (110) junctions (fig. 1) shows a similar behavior with a non-measurable ZBCP splitting and a Gap Like Feature (GLF) located at 15.8meV, the evolution of the splitting for increasing and decreasing the magnetic fields is quite different (fig. 2) for the two orientations.

In increasing fields for (100) films (fig. 1A), the splitting saturates above 1 Tesla at a value of 1.4meV. For the (110) films (fig. 1C), the splitting is much larger and does not saturate in high fields. For fields higher than 4T the splitting is smeared into the GLF structure with a maximum measurable splitting of the order of 4.5meV.

In decreasing fields, we notice that for the (100) films (fig 1B), the splitting is very small. Any splitting that is observed can be associated to some misoriented grains in the sample and the split value is within our measurement resolution. The origin for such misorientations is a direct outcome from the growing procedure. We have observed that for better (100) oriented films the splitting after field reversal almost vanishes, as can be seen in other experiments [6]. In contrast, for the (110) films (fig 1D) the splitting in decreasing fields is large and has a strong field dependency.

Following FSR model [1] and considering (100) or (103) oriented samples [6,7], it is possible to identify the increase of the splitting at low fields to the increasing screening current up to the field of the first vortex entry [2], followed by saturation [1]. The reduction in the splitting upon field reversal is due to the fast decrease in the screening currents as follows from Clem's calculation [12]. But the evolution of the splitting is peculiarly different in (110) films [8] in two major aspects. First, the splitting does not saturate at high fields and reaches values more than twice as large as any splitting seen for samples with an orientation different from (110). Second, the splitting in decreasing fields, for the (110) films, has a much stronger field dependency.

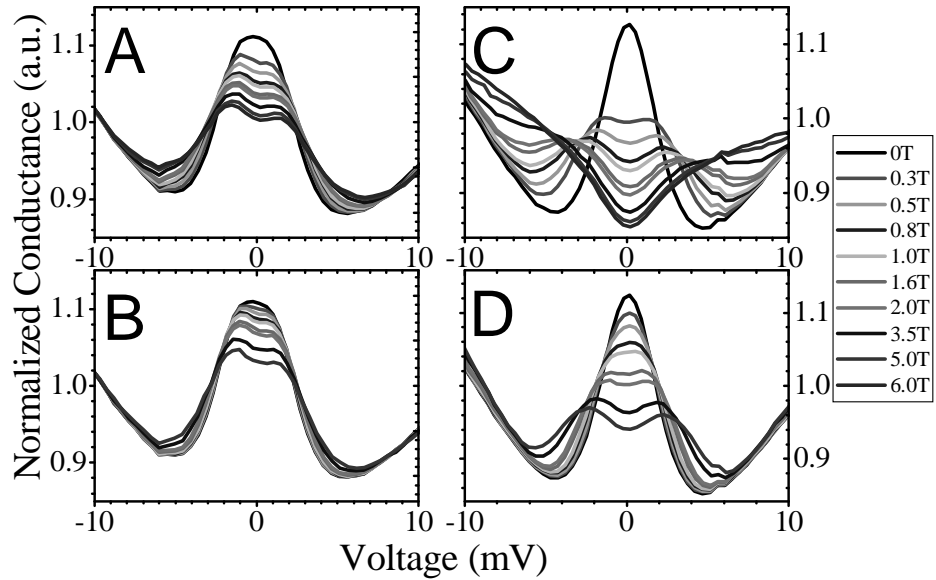

Fig. 1. Normalized *Dynamical conductance G=dI/dV Vs bias V for increasing (A,C) and decreasing (B,D) applied magnetic fields for YBCO (100) (A,B) and (110) (C,D) oriented film. Film characteristics: Tc= 89 $^oK$, GLF position 15.8mV, film thickness t=3200Å, measurement at 4.2 $^oK$. The splitting is defined as half of the distance between the positions of the conductance maxima.*

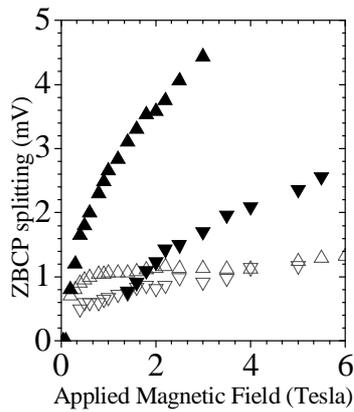

Fig. 2. *The ZBCP splitting position in increasing (triangle facing up) and decreasing fields (triangle facing down) for (110) oriented films (full triangles) and (100) oriented films (empty triangles).*

**Origin of hysteretic field splitting of the ZBCP in (100)
and (110) oriented $Y_1Ba_2Cu_3O_{7-x}$ films**

## 4. CONCLUSION

In conclusion, we have found that it is necessary to take into account the existence of a Bean-Livingston barrier in order to explain the hysteretic behavior of the ZBCP splitting in magnetic field. We have also shown that the splitting cannot be explained by the FSR model alone. The existence of a second contribution to the field splitting is necessary in order to describe the field dependency for (110) films in decreasing fields when flux can depart from the sample with no barrier. The latter contribution can be associated to the appearance of a sub dominant OP of an $id_{xy}$ symmetry which is energetically favorable as was first suggested by Laughlin [3].

## ACKNOWLEDGMENTS

We are indebted to Malcolm Beasley, Yoram Dagan and Alexander Gerber for helpful discussions. This work was supported by the Heinrich Herz-Minerva Center for High Temperature Superconductivity, by the Israel Science Foundation and by the Oren Family Chair of Experimental Solid State Physics.